\input{psfig}
\documentstyle[12pt]{article}

\title{ \normalsize  \bf PERFORMANCE AND STABILITY ANALYSIS OF A SHROUDED-FAN UAV}
\author{{ { Nicola de Divitiis}\thanks{Department of Mechanics and Aeronautics, via Eudossiana, 18, 00184
} }
 \\ {\it  University of Rome ``La Sapienza'',}
{ \it  Rome, Italy       
} 
} \date{}

\begin{document}
\baselineskip=0.900cm
\thispagestyle{empty}

\maketitle

\newcommand{\no}{\noindent}
\newcommand{\be}{\begin{equation}}
\newcommand{\ee}{\end{equation}}
\newcommand{\bea}{\begin{eqnarray}}
\newcommand{\eea}{\end{eqnarray}}
\newcommand{\bc}{\begin{center}}
\newcommand{\ec}{\end{center}}
\newcommand{\ds}{\displaystyle}

\newcommand{\bfv}{\mbox{\boldmath $v$}}

\newcommand{\bflambda}{\mbox{\boldmath $\lambda$}}
\newcommand{\bfeta}{\mbox{\boldmath $\eta$}}
\newcommand{\bfrho}{\mbox{\boldmath $\rho$}}
\newcommand{\bfomega}{\mbox{\boldmath $\omega$}}
\newcommand{\bfeps}{\mbox{\boldmath $\varepsilon$}}
\newcommand{\bfphi}{\mbox{\boldmath $\Phi$}}
\newcommand{\bfchi}{\mbox{\boldmath $\chi$}}
\newcommand{\bfxi}{\mbox{\boldmath $\xi$}}
\newcommand{\bfGamma}{\mbox{\boldmath $\Gamma$}}
\newcommand{\bfLambda}{\mbox{\boldmath $\Lambda$}}
\newcommand{\bint}{\mbox{ \int{a}{b}} }

\bc
 {\bf Abstract}
\ec

This paper deals with the estimation of the performance and
stability for a shrouded-fan unmanned rotorcraft whose mission profile
also prescribes the flight in ground effect.
The not so simple estimation of the aerodynamic coefficients and of the thrust
in the various situations makes the performance calculation and the stability analysis difficult tasks.
This is due to the strong interaction between the fan flow and shroud that causes quite different
flow structures about the airframe depending on flight conditions.
A further difficulty is related to the ground effect which produces substantial
modifications in the rotor thrust and aerodynamic coefficients.
To evaluate performance and stability, two models have been developed.
One determines the aerodynamic coefficients of the shroud, whereas
the other one calculates thrust and moment of the rotors system.
Both models take into account the mutual interference between fan flow and fuselage and ground effect.
Performance and stability are then discussed with reference to significant flight conditions.

\bc
{\bf Nomenclature}
\ec

{
\no \begin{tabular}{ll}
$C_l, C_m, C_n$                                                  & hull aerodynamic moment coefficients \\\\
$C_{l_T}, C_{m_T}, C_{n_T}$                                      & rotor  moment coefficients \\\\
$C_X, C_Y, C_Z$                                                  & hull aerodynamic force coefficients \\\\
$C_{X_T}, C_{Y_T}, C_{Z_T}$                                      & thrust force coefficients \\\\
$C_T \equiv -{C_{Z_T}}_l -{C_{Z_T}}_l$                           & rotors system  thrust coefficient \\\\
$C_Q \equiv {C_{n_T}}_l -{C_{n_T}}_u$                            & rotors system  torque coefficient \\\\
$d$                                                              & rotor diameter \\\\
$D$                                                              & overall shroud diameter   \\\\
$g$                                                              & acceleration due to gravity \\\\  
$h$                                                              & rotorcraf altitude \\\\  
$I_x, I_y, I_z$                                                  & principal moments of inertia \\\\
$I_{x z}$                                                        & product of inertia \\\\
$J$                                                              & merit function \\\\
$\ds k = \frac{V}{w_D}$                                          & velocity parameter  \\\\
$L$, $M$, $N$                                                    & hull aerodynamic moment components \\\\
$L_T$, $M_T$, $N_T$                                              & rotor moment components  \\\\
$Q = {N_T}_l -{N_T}_u$                                           & rotors system   torque \\\\
$R$                                                              & rotor radius \\\\
$S =\ds \frac{\pi {D}^2}{4}$                                     & rotorcraft reference surface \\\\
${\bf u} \equiv (\delta_A, \delta_B, \delta_C, \delta_P)$        & control vector \\\\
${\bf v} \equiv (u, v, w)$                                       & inertial velocity in body axes \\\\
${\bfv}$                                                         & local fluid velocity  \\\\
$\ds \hat {\bf v} \equiv \frac{\bf v}{V} = (\hat {u}, \hat {v}, \hat {w})$             & dimensionless velocity in body axes 
\end{tabular}}

{
\no \begin{tabular}{ll}
$V$                                                              & velocity modulus \\\\
$w_D$ = $w_i$ - $V \sin \alpha$                                  & fluid  velocity through the disk    \\\\
$W$                                                              & weight     \\\\
$w_i$                                                            & induced velocity by the rotor flow on the rotor disk \\\\
$x, y, z$                                                        &inertial coordinate \\\\
$x_B, y_B, z_B$                                                  & coordinate in body axes \\\\
$X, Y, Z$                                                        & aerodynamic force components in body axes \\\\
$X_T, Y_T, Z_T$                                                  & thrust force components in body axes\\\\
$\bf x$                                                          & position  
\end{tabular}}

\bigskip

\no {\sl Greek symbols }

{
\no \begin{tabular}{ll}
$\alpha, \beta_x$                                                & aerodynamic angles \\\\
$\alpha_x, \beta$                                                & angles of attack and sideslip, respectively \\\\
$\alpha_0 \equiv \ds \frac{\pi}{2} - \alpha_x$                     & axial angle of attack  \\\\
${\delta}_A$                                                     & lateral cyclic control, positive right  \\\\
${\delta}_B$                                                     & longitudinal cyclic control, positive after  \\\\
${\delta}_C$                                                     & collective control, positive up  \\\\
${\delta}_P$                                                     & differential collective control, positive right  \\\\
$\varepsilon$                                                    & ground inclination with respect to the horizontal plane  \\\\
$\varphi$, $\vartheta$, $\psi$                                   & Euler angles \\\\
$\bflambda \equiv (\lambda_x, \lambda_y, \lambda_z)$             & ground normal unit vector in body axes \\\\
$\gamma$ = $\ds \frac{V}{\Omega R}$                              & advance ratio  \\\\
$\gamma_x$ = $\ds \frac{u}{\Omega R}$,    $\gamma_y$ = $\ds \frac{v}{\Omega R}$,  $\gamma_z$ = $\ds \frac{w}{\Omega R}$  & dimensionless velocities    \\\\
$\lambda_i$ = $\ds \frac{w_i}{\Omega R}$                             & dimensionless induced velocity at the disk 
\end{tabular}
}

{
\no \begin{tabular}{ll}
${\Pi}_n$                                                        & required power  \\\\
$\rho$                                                           & air density \\\\
$\tau  \equiv \sqrt{1+4 (h/R)^2} -2 h/R$                         & ground effect perturbative parameter  \\\\
($p$, $q$, $r$)                                                  & angular velocity vector in body axes \\\\
$\Omega$                                                         & rotors angular velocity  \\\\
\end{tabular}
}

\bigskip

\no {\sl Subscript }

{
\no \begin{tabular}{ll}
a.c.                                                               & aerodynamic center \\\\
c.g.                                                               & vehicle center of gravity \\\\
u                                                                  & upper rotor \\\\
l                                                                  & lower rotor \\\\
$\infty$                                                           & value calculated out of ground effect
\end{tabular}
}

\bigskip

\bc
{\bf Introduction}
\ec


Performance and stability calculation of a shrouded-fan unmanned
aerial vehicle (UAV) is a very difficult task for various reasons.
One of these is related to the intensive aerodynamic interaction between shroud and rotors$^{1, 2}$
that, because of the large excursion of angle of attack and fans system induced velocity,
determines quite different flow fields about the UAV$^{3-9}$.
The simultaneous variations of angle of attack, flight speed  and the rotors induced velocity, determine
a wide set of flight conditions, where the aerodynamic and propulsive force and moment coefficients
exhibit large variations.
Furthermore the vicinity of the ground makes this study more complex, due to the
influence that the ground exerts on both fuselage and rotors system.
Therefore, the performance and stability analysis requires an accurate vehicle aerodynamics that calculates
all the aerodynamic coefficients in a wide range of angle of attack and fans working regime,
and an adequate model to determine the rotors force and moment in the different flight conditions.
The UAV here considered is a shrouded-fan made of a toroidal airframe  at
the center of which are placed two counter-rotating rotors driven by three
two-stroke air cooled engines (see Fig. 1), the main characteristics of which are reported in Tab. 1.
This vehicle is similar in shape to that studied in Ref. 1 and is the results of a project jointly accomplished
 from the University of Rome "La Sapienza"$^2$ and the Polytechnic of Turin$^1$.
The fuselage geometry is the result of a parametric study which analyzed the lift capability of
different toroidal geometries in the presence of the rotor flow.
The selected fuselage cross section develops the highest lift in hovering for an assigned hull volume.
The main function of the hull consists in to incorporate the avionics, fuel, payload and the possible
flight related hardware.
From an aerodynamic perspective, the shroud should contain the streamlines to follow the duct
and exit at an adequate velocity imposed by the exit area.
Moreover the fuselage causes an intense suction force, which is produced by the inlet lip,
that makes the fuselage a body with a nonnegligible lift capability.
Hence the shroud generates a sizable improvement of the static thrust which depends on the reciprocal
interaction between fans system and fuselage$^{5}$.
During axial flights, the advantage in static thrust furnished by the shroud rapidly diminishes$^{5}$
as soon as the UAV increases its speed.


For what concerns the ground proximity, it is responsible for several effects.
One of these is represented by the fuselage aerodynamic coefficients variations
which occur in hovering.
In such situation $\varphi = \vartheta$ = 0, then the ground imposes a boundary condition
on the aerodynamic field that symmetrically reduces the downwash of the rotors.
The consequence is that the aerodynamic coefficients depend on the dimensionless height $h/R$,
and an important loss of the fuselage lift capability is observed$^{10, 11}$.
If the vehicle flies at an arbitrary inclination with respect to the ground, the
interaction between rotors flow and ground alters the rotorcraft aerodynamics in such a way
that the pressure distribution on the airframe depends on the height and on the vehicle attitude.
Hence the aerodynamic coefficients, which are functions of the aerodynamic angles, in ground proximity also depend on the
dimensionless height and on the Euler angles.


The vicinity of the ground also influences the rotors system.
It has been observed that for an unducted rotor in hovering the thrust considerably
increases as the height diminishes$^{4-9}$.
Furthermore the attitude modifies rotor actions$^{12}$
so that, in ground effect thrust and torque depend on the Euler angles.
As a result, the attitude can play an important role for the determination of the flying qualities
depending on the height.
Also the ground inclination with respect to the horizon is of great importance
in ground effect.
This can be the case of a rotorcraft that approaches a ship flight deck$^{12, 13}$ whose inclination
influences the vehicle aerodynamics and rotors thrust and moment$^{12}$.
The deck oscillations due to the ship attitude motion determine unsteady
thrust and aerodynamic force and moment which influence the rotorcraft dynamics.
Another source of the significant variations of the performance and stability in ground effect is nonzero
flight path angle which determines incremental aerodynamic force and thrust$^{14}$.
This may happen during maneuvers that are used to avoid collisions with low obstacles.
The aforementioned effects are more pronounced at low height where the ground effect is
more intense and can determine sizable variations of trim controls and rotorcraft stability margin.

Although several studies on the mutual interference between
hull aerodynamics, rotor flow and ground effect$^{12, 14-17}$ were developed,
to the author's knowledge a general analysis of these phenomena in terms of rotorcraft
flight dynamics has not received due attention.
Therefore, the objective of the present work is to develop an accurate mathematical model that
may be adopted for the analysis of the performance and flying qualities of the UAV.

In the present study two models are proposed.
The first one calculates the shroud aerodynamic coefficients, whereas the second one determines the rotors force and
moment through the blade-element theory.
Both models account for the mutual interaction between hull and rotors and the ground proximity.
The aerodynamic model exhibits free parameters that have to be identified.
These models allow the forces and moments to be accurately formulated and the performance and flying quality to be interpreted.
Obtained results are compared with data reported in the literature,  and
finally a detailed analysis of the performance characteristics and flying qualities of the vehicle is carried out.

\bigskip

\bc
{\bf Aerodynamic Model}
\ec

A method to calculate the aerodynamic force and moment on the UAV shroud configuration, is now presented.
This procedure, based on physical considerations,  takes into account the reciprocal influence
between rotors flow and fuselage and the possible presence of the ground.
First, it is necessary to establish the flow structure about the airframe that in turn directly depends on
the rotors system working regime.
Fig. 2 shows the regions of the four characteristic working regimes in terms of the induced and
axial velocities$^{3-6}$.
In this study the vehicle aerodynamics is modelled according to the normal working state which corresponds
to situations where the induced velocity is sufficientely high with respect to the axial velocity $V \sin \alpha$.
In this situation the rotors flow enters from the top of the rotorcraft, the air is then ejected towards
the bottom and the power is transfered from the rotors to the air.
The other working regimes, such as the windmill-brake state, the turbulent wake state and the vortex ring state
are not studied in the present analysis since they are considered to be off-design situations which 
correspond to more complicate flow conditions that strongly reduce the hull lift capability.

Figure 3 shows the rotorcraft in normal working state, out of ground effect.
The rotors flow and the external stream are separated by the wake $W$ which originates from
the separation line $\Gamma$ that is assumed to be equal to the circumference of the exit area.
Because of the axial symmetry around $z_B$, the sideslip angle does not influence the shroud
aerodynamics.
Aerodynamic force ${\bf F}_A$  and moment ${\bf M}_A$ lie on the plane $z_B-V$ which is
defined by $z_B$ and ${\bf v}$.
This plane defines a reference frame ($x_v$, $y_v$, $z_v$) whose axes are oriented so as  $z_v \equiv z_B$,
$y_v$ is normal to both $z_v$ and ${\bf v}$, whereas  $x_v$ is perpendicular to both $y_v$ and $z_v$.
In this frame ${\bf F}_A$ and ${\bf M}_A$ are expressed as
\bea
 \begin{array}{l@{\hspace{0cm}}l}
{\bf F}_A =\ds   (\bar{X}, 0, \ \bar{Z}) \\\\
{\bf Q}_A = \ds  (0, \ \bar{M}, \ 0)
\end{array}
\label{aero0}
\eea
where all the components are independent on the sideslip angle.
Due to the axial symmetry, instead of using the angle of attack $\alpha_x$ and sideslip $\beta$,
it is convenient to introduce different aerodynamic angles, i.e. $\alpha$ and $\beta_x$, that are defined as follows$^{1, 2}$
\bea
\left[\begin{array}{c}
u \\\\
v \\\\
w
\end{array}\right]
\equiv V
\left[\begin{array}{c}
\hat{u} \\\\
\hat{v}  \\\\
\hat{w}
\end{array}\right]
= V
\left[\begin{array}{c}
\cos \alpha \ \cos \beta_x \\\\
\cos \alpha \ \sin \beta_x \\\\
\sin \alpha
\end{array}\right]
\eea
The components in body axes of ${\bf F}_A$ and ${\bf Q}_A$  in terms of $\beta_x$ are expressed as
\bea
\begin{array}{l@{\hspace{0cm}}l}
X =   \bar{X} \ \cos \beta_x, \ \ \
Y =   \bar{X} \ \sin \beta_x, \ \ \
Z =   \bar{Z}, \ \\\\\
L = - \bar{M} \ \sin \beta_x, \ \ \
M =   \bar{M} \ \cos \beta_x, \ \ \
N =   0
\end{array}
\eea
The hull aerodynamics is strongly influenced by the rotors flow.
In this model an equivalent actuator disk is placed in the mid of
the two rotors, whose induced velocity $w_i$ coincides with that produced by the rotors
at the actuator disk plane.
Therefore, the rotor flow is supposed to be generated by an uniform layer of doublets
with a proper value of $w_i$.
The induced velocity, that in turn represents the effect of the collective pitch on the vehicle aerodynamics,
is determined by means of the rotor model described later in the dedicated Section.
Non-uniform distributions of $w_i$ on the disk plane, caused by the cyclic pitch,
are neglected.

To derive the analytical expressions of the aerodynamic force and moment coefficients,
the rotorcraft is considered in a steady state potential flow.
Then, the flow velocity in a point $\bf x$ of the space is given by$^{18}$
\be
{\bfv}({\bf x}) = \ds \frac{\partial \bfv}{\partial \bf v}  \ {\bf v} +
 \frac{\partial \bfv}{\partial w_D}  \ w_D
\label{local_v}
\ee
where ${\partial \bfv}/{\partial \bf v}$ and ${\partial \bfv}/{\partial w_D}$ are influence matrices.
In Eq.(\ref{local_v}) the first term gives the effect of the flight velocity on the vehicle aerodynamics,
whereas the second one  provides the contribution caused by the rotors flow.
Force and moment depend on the local pressure distribution which is related to the square of the local velocity
\be
{\bfv}^2 = (\mu_1 \cos^2 \alpha + \mu_2 \sin2 \alpha + \mu_3 \sin^2 \alpha ) V^2 +
2 V w_D (\chi_1 \cos \alpha + \chi_2 \sin \alpha  ) + {g}^2 {w_D}^2
\label{2}
\ee
by means of  the Bernoulli theorem, where
$\mu_i$ and $\chi_i$ are functions of the position through ${\partial \bfv}/{\partial \bf v}$ and
${\partial \bfv}/{\partial w_D}$.
Because of the symmetry around $z_B$, the pressure does not depend on the sideslip angle.
The aerodynamic force and moment are then expressed as surface integrals of the local pressure
\be
{\bf F}_A =\int \int_S p \ {\bf n} \ dS, \ \ \ {\bf Q}_A = \int \int_S ({\bf r}-{\bf r}_{c.g.}) \times {\bf n} \ p \ dS
\label{4}
\ee
Substituting Eq. (\ref{2}) into Eqs. (\ref{4}) one obtains the aerodynamic coefficients on the plane $z_v-V$ in terms
of $k$ and $\alpha$
\bea
 \begin{array}{l@{\hspace{0cm}}l}
\ds \bar{C_X} (\alpha, k) = F_1(k, \alpha) X_1 \sin 2 \alpha + F_2(k, \alpha) X_2 \cos \alpha \\\\
\ds \bar{C_Z} (\alpha, k) = F_0(k, \alpha) Z_0 + F_1(k) Z_1 \sin \alpha +  F_2(k, \alpha) (Z_{2 1} \cos^2 \alpha + Z_{2 2} \sin^2 \alpha) \\\\
\ds \bar{C_m} (\alpha, k) = F_1(k, \alpha) M_1 \sin 2 \alpha + F_2(k, \alpha) M_2 \cos \alpha
\end{array}
\label{aero_coeff}
\eea
that are defined by means of 
\bea
 \begin{array}{l@{\hspace{0cm}}l}
{\bf F}_A =\ds \frac{1}{2} \rho V_f^2  S  \  (\bar{C}_X, 0, \ \bar{C}_Z) \\\\
{\bf Q}_A = \ds \frac{1}{2} \rho V_f^2  S D \  (0, \ \bar{C_m}, \ 0)
\end{array}
\label{aero}
\eea
where
\be
V_f = \sqrt{u^2 + v^2 + (w - w_i)^2}\equiv \sqrt{V^2 \cos^2 \alpha + w_D^2},
\ee
while the functions $F_1$, $F_2$ and $F_3$ are given by
\bea
\begin{array}{l@{\hspace{0cm}}l}
F_0 (k, \alpha) =  \ds \frac{1}  {k^2 \cos^2 \alpha + 1}, \
F_1 (k, \alpha) =  \ds \frac{k}  {k^2 \cos^2 \alpha + 1}, \
F_2 (k, \alpha) =  \ds \frac{k^2}{k^2 \cos^2 \alpha + 1}
\end{array}
\label{F123}
\eea
The aerodynamic coefficients in body axes are then obtained as the projection of Eqs. (\ref{aero_coeff})
in body frame, i.e.
\bea
\begin{array}{l@{\hspace{0cm}}l}
C_X =   \bar{C}_X \ \cos \beta_x, \ \ \
C_Y =   \bar{C}_X \ \sin \beta_x, \ \ \
C_Z =   \bar{C}_Z, \ \\\\\
C_l = - \bar{C}_m \ \sin \beta_x, \ \ \
C_m =   \bar{C}_m \ \cos \beta_x, \ \ \
C_n =   0
\end{array}
\eea
The angular velocity is taken into account
by rotational derivatives, the expressions of which are$^{18}$
\bea
\begin{array}{l@{\hspace{1cm}}l}
{C_X}_p = 0, \ \ {C_X}_q = M_{3 3} \ \sin \alpha, \ \ {C_X}_r = - M_{1 1} \cos \alpha \sin \beta_x \\\\
{C_Y}_p = -M_{3 3} \sin \alpha, \ \ {C_Y}_q = 0, \ \ {C_Y}_r = M_{1 1} \cos \alpha \cos \beta_x \\\\
{C_Z}_p = M_{1 1} \cos \alpha \sin \beta_x, \ \ {C_Z}_q =  - M_{1 1} \cos \alpha \cos \beta_x, \ \ {C_Z}_r = 0 \\\\
{C_l}_p = K_{1 2} \sin \alpha, \\\\
{C_l}_q = (K_{3 3}-K_{1 1}) \sin \alpha, \\\\
{C_l}_r = (K_{3 3}-K_{1 1}) \cos \alpha \sin \beta_x  -K_{1 2} \cos \alpha \cos \beta_x  \\\\
{C_m}_p =  (K_{1 1}-K_{3 3}) \sin \alpha \\\\
{C_m}_q = K_{1 2} \sin \alpha, \\\\
{C_m}_r = (K_{1 1}-K_{3 3}) \cos \alpha \cos \beta_x  -K_{1 2} \cos \alpha \sin \beta_x
\end{array}
\label{rotary}
\eea
where
\bea
\begin{array}{l@{\hspace{1cm}}l}
M_{i  j} = M_{i  j \ 0} F_0 (k, \alpha)   + M_{i  j \ 1} F_1 (k, \alpha)  \\\\
K_{i  j} = K_{i  j \ 0} F_0 (k, \alpha)   + K_{i  j \ 1} F_1 (k, \alpha)
\end{array}
\label{rotary1}
\eea
Eqs. (\ref{rotary}) represent the expressions of the rotational derivatives of a body which exhibits an axial symmetry around $z_B$.
The quantities $M_i$, $X_i$, $Z_i$ appearing in Eqs. (\ref{aero_coeff}) and  $Z_{i j}$, $M_{i j \ 0}$,  $M_{i j \ 1}$, $K_{i j \ 0}$ and  $K_{i j \ 1}$ appearing in Eqs.(\ref{rotary}- \ref{rotary1}) are free parameters which depend on the vehicle geometry.

\bigskip

\bigskip


In ground effect, the rotor downwash is limited by the ground surface,
so that the velocity field about the airframe is modified with respect to Eq. (\ref{local_v}).
To study this situation consider now Fig. 4, where the rotorcraft in close proximity to the flat ground is analyzed.
The wake $W$ is considered to be a rigid cylinder whose length is assigned in advance in such a way that it
does not intersect the ground plane.
The ground inclination with respect to the rotorcraft is represented by the unit vector
\be
{\bflambda} \equiv (\lambda_x, \lambda_y, \lambda_z) = (- \sin \vartheta, \sin \varphi \cos \vartheta, \cos \varphi \cos \vartheta)
\ee
which is the normal unit vector of the ground plane in body axes.
The presence of the ground is simulated by placing a specular image of the rotorcraft at an
equal distance below the ground plane.
According to the potential flow theory$^{18}$, the local velocity is the sum of those induced by the vehicle
and by its mirror image, i.e.
\be
\begin{array}{l@{\hspace{0cm}}l}
{\bfv}({\bf x}) =
{\bfv}_{\infty}({\bf x}) +
 \ds \frac{\partial {\bfv} }{\partial \{ {\bflambda} w_D \}} w_D {\bflambda} +
 \ds \frac{\partial^2 {\bfv} }{\partial {\bflambda} \partial{\bf v}}  {\bflambda}  {\bf v}
\end{array}
\label{1a}
\ee
The first addend is the local velocity induced by the entire rotorcraft, whereas the second and the third ones provide, respectively,
the interaction between rotors flow and ground plane and the influence that the ground exerts on the fuselage.
The pressure about the airframe is then altered by the vicinity of the ground,
so that the aerodynamic coefficients also depend on height and attitude.
Now, to obtain the expressions of the force and moment in ground effect,
the kinetic energy of the stream $T$ is considered.
Following  Ref. 18, $T$ is a definite positive quadratic form that is a function of  $\bf v$, $w_D$ and $\bflambda$.
In order to determine the expression of $T$, consider now two sets of flight conditions.
The first of these corresponds to horizontal flights
[ $ \bflambda=(0, 0, 1)$ ], where  $\bf v$ describes a circular cone about $z_B$, whereas
the second one consists of axial flights [ $\bf v$ = (0, 0, V) ] with
$\bflambda$ which makes a circular cone about $z_B$.
In both cases, due to the axial symmetry about $z_B$, the kinetic energy must remain unaltered.
As a result the kinetic energy of the stream can be written in the form
\be
T  =  f_1 (u^2+ v^2)  + f_2 w^2 + f_3 \ w_D (u \lambda_x + v \lambda_y)  + f_4 \ w_D  w \lambda_z
\label{kinetic}
\ee
where $f_1$, $f_2$, $f_3$ and $f_4$ are functions of $h/R$ that,
according to Ref. 19, are written in terms of $h/R$ by means of the perturbative parameter $\tau$, i.e.
\be
\ds f_k =  \sum_{m = 1}^3 f_{k m} \tau^m, \ \ \ (k=1, 2, 3, 4)
\label{perturbative_parameter}
\ee
The shroud aerodynamic force and moment increments developed in ground effect
are then expressed using the Lagrange equations method$^{18}$
\bea
\begin{array}{l@{\hspace{0cm}}l}
\ds \Delta {\bf F} \equiv
(\Delta X,  \ \Delta Y, \ \Delta Z )
= - \frac{\partial }{\partial t} \frac{\partial T}{\partial {\bf v}}
  -\frac{\partial }{\partial h} \frac{\partial T}{\partial {\bf v}} \ V \sin \gamma, \\\\
\ds \Delta {\bf Q} \equiv
(\Delta L,  \ \Delta M, \ \Delta N )
= -{\bf v} \times  \frac{\partial T}{\partial {\bf v}}  + {\bf r}_{AG} \times {\Delta \bf F}
\end{array}
\label{force_moment}
\eea
where ${\bf r}_{AG} \equiv {\bf r}_{a.c.}-{\bf r}_{c.g.}$ = (x$_{AG}$, y$_{AG}$, 0) is the position of the
aerodynamic center with respect to c.g., which varies in ground effect.
Out of ground effect it is assumed that ${\bf r}_{a.c.} = {\bf r}_{c.g.}$ while
close to the ground, ${\bf r}_{a.c.}$ depends upon the flight conditions.
Since the yaw moment must be equal to zero in all the situations, the projection along
$z_B$ of ${\bf r}_{AG} \times {\Delta \bf F}$  must be identically equal to zero.
The consequence is that ${\bf r}_{AG}$ assumes the form
\be
\ds \frac{{\bf r}_{AG}}{R} =  \zeta  ( \frac{\Delta X}{Z_\infty}, \ \frac{\Delta Y}{Z_\infty}, \ 0 \ )
\ee
where $\zeta$ is a quantity depending on $h/R$ that is expressed through $\tau$
\be
\ds \zeta =  \sum_{k = 0}^3 \chi_k \tau^k
\label{chi}
\ee
Substituting Eq. (\ref{kinetic}) into Eqs.(\ref{force_moment})
one obtains the incremental force and moment in ground effect
\bea
\begin{array}{l@{\hspace{0cm}}l}
\ds \Delta { X} = -( 2 \dot{f}_1 u + \dot{f}_3 w_D \lambda_x  )-( 2 {f_1}_h u + {f_3}_h w_D \lambda_x) \ V \sin \gamma \\\\
\ds \Delta { Y} = -( 2 \dot{f}_1 v + \dot{f}_3 w_D \lambda_y  )-( 2 {f_1}_h v + {f_3}_h w_D \lambda_y) \ V \sin \gamma \\\\
\ds \Delta { Z} = -( 2 \dot{f}_2 w + \dot{f}_4 w_D \lambda_z  )-( 2 {f_2}_h w + {f_4}_h w_D \lambda_z) \ V \sin \gamma
\end{array}
\label{force_lagr}
\eea
\bea
\begin{array}{l@{\hspace{0cm}}l}
\ds \Delta L = -2 ({f}_2-{f}_1) \ v w - {f}_3 w_D  \lambda_z v-
\chi R [ (2  \dot{f}_1 v + \dot{f}_3 w_D \lambda_y) + ( 2 {f_1}_h v + {f_3}_h w_D \lambda_y) \ V \sin \gamma] \\\\
\ds \Delta M = 2 ({f}_2-{f}_1) \ u w + {f}_3 w_D  \lambda_z u
+\chi R [(2  \dot{f}_1 u + \dot{f}_3 w_D \lambda_x) + ( 2 {f_1}_h u + {f_3}_h w_D \lambda_x) \ V \sin \gamma] \\\\
\ds \Delta {N} = 0
\end{array}
\label{moment_lagr}
\eea
where the quantities $\dot{f}_k$ and $f_h$ are
\be
\ds \dot{f}_k = \dot{f}_{k V} V + \dot{f}_{k w_D} w_D, \ \ {f_k}_h = f_{k V} V + f_{k w_D} w_D
\ee
Following the classical theory of the lifting bodies,  both $\dot{f}_k$ and $f_h$ are proportional to the
circulation around the hull.
In particular, $\dot{f}_k$ is the time derivative of $f_k$, which is caused by the wake,
while ${f_k}_h$ is the derivative of $f_k$ with respect to $h$ which is related to the sink rate.
Accounting for the variations of $\zeta$ and $f_k$, the aerodynamic force coefficients can be written in the form
\bea
\begin{array}{l@{\hspace{0cm}}l}
\ds \Delta {C_x} =  \sum_{m = 0}^2  \sum_{n = 1}^3 F_m(k, \alpha) \tau^n
[(A_{m n} \hat{u} + B_{m n} \lambda_x ) + (C_{m n} \hat{u} + D_{m n} \lambda_x ) \sin \gamma]   \\\\
\ds \Delta {C_y} =  \sum_{m = 0}^2  \sum_{n = 1}^3 F_m(k, \alpha) \tau^n
[(A_{m n} \hat{v} + B_{m n} \lambda_y ) + (C_{m n} \hat{v} + D_{m n} \lambda_y ) \sin \gamma]   \\\\
\ds \Delta {C_z} =  \sum_{m = 0}^2  \sum_{n = 1}^3 F_m(k, \alpha) \tau^n
[(E_{m n} \hat{w} + G_{m n} \lambda_z ) + (H_{m n} \hat{w} + L_{m n} \lambda_z ) \sin \gamma]
\end{array}
\label{cforce_lagr}
\eea
whereas the aerodynamic moment coefficients are
\bea
\begin{array}{l@{\hspace{0cm}}l}
\ds \Delta C_l = \sum_{m = 0}^2 \sum_{n = 1}^3 F_m(k, \alpha) \tau^n
[(N_{m n} \hat{v} + P_{m n} \hat{v} \hat{w} + Q_{m n} \lambda_z \hat{v} + R_{m n} \lambda_y  ) + \\
 \hspace{47.mm} (S_{m n} \hat{v} + T_{m n} \hat{v} \hat{w} + U_{m n} \lambda_z \hat{v} + V_{m n} \lambda_y  ) \sin \gamma]\\\\
\ds \Delta C_m = -\sum_{m = 0}^2 \sum_{n = 1}^3 F_m(k, \alpha) \tau^n
 [(N_{m n} \hat{u} + P_{m n} \hat{u} \hat{w} + Q_{m n} \lambda_z \hat{u} +R_{m n} \lambda_x  ) + \\
  \hspace{52.mm} (S_{m n} \hat{u} + T_{m n} \hat{u} \hat{w} + U_{m n} \lambda_z \hat{u} + V_{m n} \lambda_x  ) \sin \gamma]\\\\
\ds \Delta C_n = 0
\end{array}
\label{cmoment_lagr}
\eea
In Eqs. (\ref{cforce_lagr}) and (\ref{cmoment_lagr}) the coefficients
$A_{m n}$, $B_{m n}$, $C_{m n}$, $D_{m n}$, $E_{m n}$, $G_{m n}$, $H_{m n}$, $L_{m n}$, $N_{m n}$, $P_{m n}$, $Q_{m n}$,
$R_{m n}$, $S_{m n}$, $T_{m n}$, $U_{m n}$ and  $V_{m n}$ are the free parameters of the aerodynamic model in ground effect.
Since for
$k \rightarrow 0$, $\Delta C_x$ = $\Delta C_y$ = $\Delta C_l$ = $\Delta C_m$ do not depend on the aerodynamic angles, one has
$A_{0 n}$ = $C_{0 n}$ = $E_{0 n}$ = $H_{0 n}$ = $N_{0 n}$ = $P_{0 n}$ = $Q_{0 n}$ = $S_{0 n}$ = $T_{0 n}$ = $U_{0 n}$ =0.
The terms with $\sin \gamma$ yield the influence of the sink rate on the shroud aerodynamics.
A non zero sink rate determines an unsteady ground effect that modifies the aerodynamic coefficients according
to the Bernoulli theorem
\be
\ds \frac{\partial \phi}{\partial t} + \frac{ {\bfv} \cdot {\bfv} }{2} + \frac{p}{\rho} = const
\ee
where ${\partial \phi}/{\partial t} =  {\partial \phi}/{\partial h} \  \dot{h}$ $\equiv {\partial \phi}/{\partial h} \   V \sin \gamma$
is the unsteady term that produces the dynamic ground effect.
As for the influence of the spin rate,  Eqs.(\ref{rotary}) are considered to be valid also in ground effect.
\bigskip

\bc
{\bf  Rotors System Model}
\ec

The rotors actions provide lift force and control moments to manage rotorcraft attitude. Pitch and roll are controlled through longitudinal $\delta_B$ and lateral $\delta_A$ variations of blade pitch, whereas the yaw control is carried out by means of differential variation $\delta_P$ of the collective pitch on both rotors, whose angular velocity is kept constant by a RPM governor. The blade pitch is controlled by a mechanism consisting of two independent swash-plates, each driven by three actuators. 

Thrust forces and moments developed by the rotors are given by the equations$^{6}$
\bea
 \begin{array}{l@{\hspace{1cm}}l}
(X_T, Y_T, Z_T) \ = - \pi \ \rho \ \Omega^2 \ R^4 \ (C_{x_T}, C_{y_T}, C_{z_T})
\\\\
(L_T, M_T, N_T) \ = \ \pi \ \rho \ \Omega^2 \ R^5 \ (C_{l_T}, C_{m_T}, C_{n_T})
\end{array}
\label{1b}
\eea
where $C_{x_T}, C_{y_T}, C_{z_T}$ and $C_{l_T}, C_{m_T}, C_{n_T}$ are, respectively, thrust and torque coefficients.
They are obtained with the blade-element theory which calculates thrust and moment through analytical integration of the aerodynamic load along the blade span assuming steady-state aerodynamics$^{1}$, where the effects of the blade-tip losses and the mutual influence between the two rotors are neglected.
The determination of the rotors  regime is made by imposing that the thrust coefficient
$C_{z_T} (\gamma_x, \gamma_y, \gamma_z)$, calculated with the blade-element theory,
is equal to that obtained by the actuator disk theory$^{20}$, that is
\be
\ds C_{z_T a} = 2 \lambda_i \sqrt{\gamma_z^2 + \lambda_i^2 - 2 \gamma_z \  \lambda_i \sin \alpha}
\ee
 where
\be
\gamma_x = \ds \frac{u+u_s}{\Omega R}, \  \gamma_y = \ds \frac{v+v_s}{\Omega R}, \  \gamma_z = \ds \frac{w+w_s}{\Omega R},
 \ds  \lambda_i = \frac{w_i}{\Omega R}
\ee
and
\be
(u_s, v_s, w_s) = \ds \frac{\partial \bfv}{\partial {\bf v}} \cdot {\bf v}
\ee
is the velocity induced by the shroud at the center of the actuator disk plane.
As a result, one obtains the following equation
\be
\ds  C_{z_T} (\gamma_x, \gamma_y, \gamma_z) = 2 \lambda_i \sqrt{\gamma_z^2 + \lambda_i^2 - 2 \gamma_z  \ \lambda_i \sin \alpha}
\label{regime}
\ee
the solution of which gives the dimensionless induced velocity $\lambda_i$ and the corresponding thrust coefficient.
To take into account the ground effect, the ground is modelled by means of the mirror image of the entire rotorcraft.
Equation (\ref{regime}) is still considered valid even though $(u_s, v_s, w_s)$ now accounts for
the induced velocity of the mirror image calculated in the origin of the actuator disk plane, i.e.
\be
(u_s, v_s, w_s) =  \ds \frac{\partial \bfv}{\partial {\bf v}} \cdot {\bf v} +
                   \ds \frac{\partial {\bfv} }{\partial {\bflambda} w_D} w_D {\bflambda} +
                   \ds \frac{\partial^2 {\bfv} }{\partial {\bflambda} \partial{\bf v}}  {\bflambda}  {\bf v}
\ee
A nonzero sink rate determines an unsteady ground effect for the rotors system, then
following the Bernoulli theorem, the pressure distribution  along each blade is augmented by the quantity
\be
\Delta p = -  \rho \frac{\partial \phi}{\partial h} V \sin \gamma
\ee
that expresses the pressure increment caused by the dynamic ground effect.

\bigskip

\bc
{\bf The equations of motion}
\ec

A full nonlinear six-degree-of-freedom model of the vehicle is now defined.
Following Etkin$^{21}$, the equations of motion  for the rotorcraft are written as follows
\bea
 \begin{array}{l@{\hspace{1cm}}l}
\ds \dot{u} = \ds g \frac{X_T + X}{W} -g \sin  \vartheta                -q \ w +r \ v
\\\\
\ds \dot{v} = \ds g \frac{Y_T + Y}{W} +g \cos  \vartheta \ \sin \varphi -r \ u +p \ w
\\\\
\ds \dot{w} = \ds g \frac{Z_T + Z}{W} +g \cos  \vartheta \ \cos \varphi -p \ v +q \ u
\\\\
\ds \dot{\varphi} = p + \sin \varphi \tan \vartheta \ q + \cos \varphi \tan \vartheta \ r
\\\\
\ds \dot{\vartheta}= \cos \varphi \ q - \sin \varphi \ r
\\\\
\ds \dot{\psi}= \sin \varphi \sec \vartheta \ q + \cos \varphi \sec \vartheta \ r
\\\\
\ds \dot{p}= \frac{[I_{x z} p q +(I_y-I_z) q r + L + L_T] I_z+  [-I_{x z} q r  +(I_x-I_y) p q + N + N_T ] I_{x z} }{I_x I_z -I_{x z}^2}
\\\\
\ds \dot{q}= \frac{I_{x z} (r^2 - p^2) +(I_z-I_x) p r + M + M_T}{I_y}
\\\\
\ds \dot{r}= \frac{[-I_{x z} q r +(I_x-I_y) p q + N +N_T] I_x +  [I_{x z} p q  + (I_y-I_z) q r + L +L_T ] I_{x z} }{I_x I_z -I_{x z}^2}
\\\\
\ds \dot{x} = (\cos \vartheta \cos \psi) \ u + (\sin \varphi \sin \vartheta \cos \psi- \cos \varphi \sin \psi) \  v
\\\
\ \ \ \ + (\cos \varphi \sin \vartheta \cos \psi + \sin \varphi \sin \psi) \ w
\\\\
\ds \dot{y} =  (\cos \vartheta \sin \psi) \ u + (\sin \varphi \sin \vartheta \sin \psi+ \cos \varphi \cos \psi) \ v
\\\
\ \ \ \ + (\cos \varphi \sin \vartheta \sin \psi - \sin \varphi \cos \psi) \ w
\\\\
\ds \dot{z} = (- \sin \vartheta) \ u + (\sin \varphi \cos \vartheta) \ v + (\cos \varphi \cos \vartheta) \ w
\end{array}
  \label{motion equations}
\eea
It is worth to remark that both the coaxial rotors cause moments that are transmitted to the fuselage. Since the rigid rotors have the same moments of inertia with respect to the rotation axis, their gyroscopic effects are balanced each other and therefore do not appear in the rigid body moment equations.

\bigskip

\bc
{\bf Identification of the Aerodynamic Model Parameters}
\ec

This section describes the identification procedure for the free parameters of the aerodynamic model.
The algorithm consists of an optimization method, based on
the least-square procedure, that is represented by the problem
\be
J =   \sum_k [(C_k)_{CFD}-C_k]^2 \ = \  min
\label{10}
\ee
where $C_k$ is the generic aerodynamic coefficient calculated with the proposed model,
whereas  $(C_k)_{CFD}$ is the same coefficient computed by CFD simulations.
These simulations are obtained using VSAERO by Analytical Methods, Inc.$^{22}$, which is a code based on a boundary integral formulation. 
The code accounts for the aforementioned effects such as the interference between fan flow and airframe, the ground effect and the
presence of the separated flows enclosed by the wake.

The design variables of the problem given by Eq. (\ref{10}) are defined as the arguments of $J$, that are the aerodynamic model free parameters.

The free parameters of the model out of ground effect are determined first. They are the coefficients
of the  Eqs. (\ref{aero_coeff}) and  (\ref{rotary1}) which are identified through several CFD calculations.
Once the simulations are carried out, the quantities
${\partial \bfv}/{\partial {\bf v}} $ and $ {\partial {\bfv} }/{\partial  w_D}$ are also determined.
The results are shown in Fig. 5, that reports drag, lift and pitch moment coefficients, where
the continuous lines and the solid symbols are, respectively, the results calculated by the model and the data obtained by VSAERO.

For what concerns the ground effect, the free parameters are the coefficients of Eqs. (\ref{cforce_lagr}) and (\ref{cmoment_lagr}).
To assess the simultaneous influence of $\alpha$, $\beta_x$, $\varphi$, $\vartheta$, $h/R$ and $k$ on the vehicle
aerodynamics in ground effect, a large number of CFD simulations is made with VSAERO, for several flight conditions.
The model free parameters are then calculated using the aforementioned minimization algorithm, so that also
the influence functions ${\partial {\bfv} }/{\partial {\bflambda} w_D}$, ${\partial^2 {\bfv} }/{\partial {\bflambda} \partial{\bf v}}$
and ${\partial \phi }/{\partial h}$ are evaluated.

\bigskip

\bc
{\bf Validation of the Models}
\ec

To validate the proposed models, the present results are compared with some existing data in the literature.

In Fig. 6 the aerodynamic coefficients in body axes in terms of $\alpha_0$ are depicted.
The present results (continuous lines) are compared with those from Ref. 23 (solid symbols). This latter model is based upon
Fourier expansion of the vorticity distribution that accounts for the effects of camber, taper and thickness.
These data are determined  by combining all these effects to obtain a similar toroidal fuselage to the shroud here studied.
Some discrepancy is apparent, especially for $C_z$ when $k$ tends to zero, that could be caused by the difference
between the two geometries.
In spite of that, the present results can be considered in somewhat comparable with those
obtained using the model of Ref. 23.

As for the ground effect, it is known that for a platform in the presence of a lifting jet$^{10}$,
the hull exhibits a substantial lift reduction$^{10, 11}$ that becomes more influent
as the fuselage approaches to the ground.  In the case of hovering (V = $\varphi$ = $\vartheta$ =0),
it can be shown through the Buckingham theorem that, for given thrust and hull geometry, the dimensionless parameters
that describe the lift loss phenomenon are $\ds \frac{\Delta Z} {Z_\infty}$ and $h/(D-d)$ $^{10, 11}$.
Fig. 7a shows the present results (continuous line) in comparison with the data from Ref. 10 and 11
(dashed line and solid symbols, respectively), that are static measurements of the lift losses in ground effect for
a circular platform.
It is worth to remark that the present data are referred to a toroidal fuselage, while Refs. 10 and 11 deals
with circular planforms with a centrally-located lifting jet. Nevertheless, the obtained results
are qualitatively in good agreement with the those of Refs. 10 and 11.

The aerodynamic coefficients in ground effect also depend on the vehicle attitude and sink rate.
Then, Fig. 7b, 7c and 7d show the increments of the longitudinal aerodynamic coefficients (continuous lines),
calculated at $\varphi =0$, $k=0$ for various $h/R$, while Fig. 7e gives the derivative
$\partial C_z / \partial \gamma$ at $\alpha = \pi/2$, v.s. $k$, where each curve represents a given $h/R$.
The results are represented together with those of a rigid vortex ring in the vicinity of a wall (solid symbols), where
the vortex geometry is assumed to be equal to the circumference defined by the centers of the shroud cross sections,
whereas the vortex circulation is selected in such a way that the induced velocity at the center of the vortex ring is equal to $w_i$.
As well as in the cases of the hull and rotors system, the wall is simulated by means of the vortex ring mirror image, where
the induced velocity on each vortex element is calculated by means of the Biot Savard law,
whereas the corresponding action is determined using the Kutta-Joukowsky theorem.
Some disagreement is evident, due to the geometrical differences between toroidal fuselage and vortex ring,
that becomes more significant as $h/R$ goes to zero.
Nevertheless, the obtained results seem to be in adequate agreement with those of the vortex ring.
Because of the symmetry around $z_B$,  $\Delta C_x$ and $\Delta C_m$ are odd functions, while $\Delta C_z$ is an even function
of $\vartheta$. $\Delta C_z$ varies in accordance to Fig. 7c and presents minor variations with $\vartheta$ but for $h/R \rightarrow$ 0,
while $C_m$ exhibits a negative slope whose absolute value increases as $h/R$ approaches to zero. This last characteristic corresponds to a
positive contribution to the rotorcraft stability in ground effect.
The plot of Fig. 7e, which  illustrates the effect of the sink rate on the hull aerodynamics, 
shows the derivative $\partial C_z / \partial \gamma$  in terms of k. A negative flight path angle
(positive sink rate) produces a negative variation of $C_z$ that corresponds to a an increment of the shroud lift coefficient.
For each curve, $\partial C_z / \partial \gamma$ exhibits its maximum value at about k=1.
As for the lateral coefficients, because of the hull symmetry, when $\vartheta$ is changed with $\varphi$,
then $C_x$, $C_m$  are changed in $C_y$ and -$C_l$, respectively.

For what concerns the rotors model, various calculations have been carried out.
Fig. 8 gives $C_T$ and $C_Q$ of the rotors system in axial flight
with (continuous lines) and without (dashed lines) shroud in terms of the advance ratio
for several collective pitch. In both cases, the rotors system exhibits
the behavior of a propeller in axial flight, whereas the presence of the  shroud produces sizable variations in the thrust.
For each collective pitch, at $\gamma =0$ (static case), the fuselage generates a thrust gain
of about 25$\%$ with respect to the unshrouded rotor, whereas, as soon as the advance ratio increases, a more
limited thrust gain is observed. As for the torque coefficient, minor variations are observed.
This results are in accordance with Refs. 3 and 5, where similar configurations of ducted propeller are treated.

The vicinity of the ground significantly modifies the rotors characteristics.
The plots in Fig. 9 show the thrust coefficient,  induced velocity and induced torque coefficient v.s. $h/R$.
In Fig. 9a the solid symbols are from Ref. 4, which represent flight tests accomplished
with different helicopters, whereas the continuous line yields the present results
that are obtained by applying the proposed rotor model to a single free rotor.
Fig. 9a also shows the Cheesman and Bennett results$^{16}$ (dot-dashed line), wherein the presence of the
ground is modelled by placing under the ground a mirror image which consists of a simple fluid source
whose mass flow coincides with that of the rotor.
The dashed lines give the data obtained using the Hayden$^{17}$ method which
estimates the influence of the ground in hovering through flight test measurements.
According to the literature$^{4, 5}$, the Hayden results are found to overpredict the rotor thrust.
The figure shows that the present results are in good agreement with the aforementioned flight tests measurements.
Fig. 9b and 9c compare the induced velocity and the torque coefficient with the other data sources.
The solid symbols in Fig. 9b are from Ref. 12, that represent the normalized average inflow in terms of $h/R$.
According to Xin and Prasad$^{12}$, the continuous line is calculated by applying the proposed model to rotor of the Yamaha R-50,
which is a small size, remotely piloted helicopter with a rotor diameter and a rotor speed of 3.07 m and 850 R.P.M., respectively.
The present results, which are compared with Ref. 12 (solid symbols, see figure), shows that the rotor model gives results 
in good agreement with those reported in the literature.
The continuous line with the filled symbols provides the data for the UAV here studied.

Also the attitude influences the rotor forces and moments in ground effect.
In the diagrams of Figs. 10a, 10b  and 10c, the coefficients $C_{Z_T}$, $C_{X_T}$ and $C_{m_T}$ at V=0 are reported
as functions of $\vartheta$, where continuous an dashed lines represent the data for shrouded and free rotors, respectively. 
As well as in the case of the hull, because of the vehicle axial symmetry, $C_{Z_T}$ is an even function, while $C_{X_T}$ and $C_{m_T}$ 
are both odd functions of the pitch angle. While the pitch angle produces minor variations on $C_{Z_T}$,
$C_{m_T}$ exhibits negative slopes whose absolute value increases as $h/R$ tends to zero. This last
characteristic is the contribution of the fans system to the vehicle stability in ground effect.
The presence of the shroud significantly modifies the thrust force and moment coefficients.
In particular a sizable $C_{Z_T}$ increment occurs, which in turn is in agreement with Fig. 8,
and an important slope reduction of both $C_{X_T}$ and $C_{m_T}$ is observed that is caused by the shroud induced velocity.

Again, following Ref. 12, the present model is now applied to the Yamaha R-50 rotor (Fig. 10 d).
The curves give the torque coefficient vs. $h/R$ at different $\vartheta$,
where the solid symbols are from Ref. 12, while the continuous lines correspond to the present results.
According to the literature$^{12}$, for each $h/R$, the vehicle attitude tends to reduce the torque coefficient, and 
the comparison with Ref. 12 shows that the maximum difference between the two methods results to be always less than 7$\%$.

Finally, to assess the dynamic ground effect of the rotors system, the thrust variations
in term of the dimensionless sink rate are shown in Fig. 10e.
Continuous and dashed lines give, respectively, the present model and the results obtained with the unsteady actuator disk theory$^{20}$, whereas the symbols represent the thrust calculated by a code based on the vortex lattice method$^{24}$, that simulates the dynamic ground effect by means of the mirror image of the actuator disk that moves with a flight path angle equal to -$\gamma$.
These comparison demonstrates the good agreement between the present data and the results obtained by the actuator disk theory and by the CFD simulations.

\bc
{\bf Results and Discussion}
\ec


In this section, to analyze the rotorcraft performance and stability in and out of ground effect,
some significant situations which correspond to straight and level flight at the trim are studied.
Eqs. (\ref{motion equations}) are used as equations of motion and the contribution of the control variables
to vehicle dynamics is taken into account through the rotors model.
The trim calculation is made by solving a minimization problem with assigned constraints$^{25}$, whereas 
the vehicle stability is investigated by means of the eigenvalues analysis applied to the linearized motion equations.

Fig. 11 depicts some of the significant variables calculated at trim.
In Fig. 11a lift  and drag coefficients are shown in terms of forward speed.
Due to the hull lift cabability caused by the suction force developed by the inlet lip,
a sizable nonzero lift coefficient is observed at low velocities, which is about 25 $\%$ greater
than that of the UAV analyzed in Ref. 1.
The diagram in Fig. 11b shows the hull trim moment as the function of the flight speed.
This is important for the longitudinal stability and controllability, since the aerodynamic moment at the trim is balanced
by the rotors control moment, generated by longitudinal cyclic pitch.
The present results calculated out of ground effect (dashed lines) can be compared with the data from Refs. 1 and 26. 
In Ref. 1, where a quite similar rotorcraft is studied, the trim moment is a rising function of the forward speed in the speed range
$0 \div 30$ m s$^{-1}$, and the moments calculated for the velocities of 10 and 20 m s$^{-1}$, are about 100 and 180 N m, respectively.
Ref. 26 deals with the Cypher, an uninhabited rotorcraft developed by 
Sikorsky Aircraft Corporation which is also made by a toroidal fuselage with at the center two coaxial rotors.
There, the aerodynamic moment rises until to a speed of about 13 m s$^{-1}$, where exhibits its maximum value of about 200 N m.
Although the two vehicles treated in Refs. 1 and 26 exhibit differences with respect to the present one,
the corresponding data seem to be in somewhat comparable with the results shown in Fig. 11b.

As seen, the ground proximity causes significant variations on the vehicle aerodynamics and on the fans regime in such a way that
both aerodynamic and thrust coefficients also vary with height and attitude.
This effect results to be more pronounced at lower $h/R$.
According to the lift reduction for platforms with lifting jet$^{10}$, 
the fuselage lift capability diminishes as the rotorcraft approaches to the ground.
The plots in Figs. 11c, 11d, 11e and 11f show  the controls expressed in radians in terms of forward speed.
For each $h/R$, the collective pitch $\delta_C$ (Fig. 11c) presents a low speed region where it remains practically constant
until to the speed of about 20 m s$^{-1}$, over which it becomes a rising function of $V$, whereas for what concerns   
the longitudinal cyclic pitch $\delta_B$ (Fig. 11d), its values are directly related to the moment calculated at the trim.
For $h/R$= 2, 3, $\delta_C$ varies in accordance to the rotors thrust increments,
whereas $\delta_B$ diminishes until to 20 m s$^{-1}$, where reaches  
its minimum. At higher velocities, according to the hull moment developed at the trim, $\delta_B$ rises with $V$.
The calculation made at $h/R$ = 1 shows that $\delta_B$ is a decreasing function of $V$ in the entire speed range, and
for what concerns $\delta_A$ and $\delta_P$, quite small variations are obtained, but for $h/R$ $<$ 1.
In this last situation the simultaneous effects of pitch angle and height generate a
strong flowfield perturbation on the two rotors in different fashions which must be balanced by
the differential collective pitch.
At $h/R$ = 1 the peripheral parts of the airframe can touch the ground for certain pitch angles,
thus the trim results to be feasible in a more limited speed range. Therefore the corresponding curves 
are broken at a flight speed less than 24 m s$^{-1}$, where $\vartheta \simeq$ 35$^o$.

For each  $h/R$, the law $\vartheta$($V$) directly depends on the forces calculated at the trim (see Fig. 11 g).
Because of the smaller variations of lift and drag with respect to 
the thrust, the pitch angle exhibits relatively small variations as $h/R$ changes.
The required power $\Pi_n$ and the velocity parameter $k$ are represented in Fig. 11h in  terms of flight speed.
Some comparisons can be made out of ground effect (dashed lines) with the data from Ref. 1.
Close the hovering, $\Pi_n$ is rather similar to that calculated in Ref. 1, whereas
for $V \ne$ 0 some discrepancies are apparent which are caused by the different fuselage geometries that develop diverse
aerodynamic actions. The higher shroud lift capability than that of the vehicle dealt in Ref. 1,  yields 
smaller required power when $V \ne$ 0.
The ground effect causes important modifications on $\Pi_n$ and $k$.
At $h/R$ = 1,  for $V < $ 10 m s$^{-1}$, a reduction of the 10 $\%$ of $\Pi_n$ is observed,
while at higher velocities, where $\alpha$ $\approx$ $\vartheta$ $\ne$ 0,
the interaction between the rotorcraft and ground generates seeable variations on both aerodynamic and thrust coefficients
which in turn are responsible for the $\Pi_n$ variations for $V$ $\ge$ 10 m s$^{-1}$.
The law k($V$) calculated out of ground effect (dashed line), is similar in shape to that of Ref. 1 where k is
about 10 $\%$ greater than that here calculated.  The downwash reduction due to the ground effect, causes
smaller values of k as $h/R$ approaches to zero.

Next, to asses the influence of the ground inclination on the trim, the calculations
are repeated in the cases where the ground wall is inclined with respect to the horizon.
The wall inclination, which is here obtained by rotating the ground surface of an angle $\varepsilon$ around the x axis,
is represented by the wall normal unit vector, that in the inertial frame is given by
\be
{\bflambda} =(0, \sin \varepsilon, \cos \varepsilon)
\ee
To avoid possible touchdown, the trim, whose results are summarized in Fig. 12, is made at $h/R$=1 for a ground inclination of
$\varepsilon =$ $\pm$  20$^o$.
In such situation the peripheral parts of the airframe are very close to the wall, thus the interaction
between rotorcraft and ground determines high rotors and fuselage moments that yield not feasible trim for V$\leq$ 20  m s$^{-1}$.
Although the state variables result to be rather similar than those just depicted in Fig. 11,
the controls are quite different with respect to those obtained for $\varepsilon$ =0.
Relevant variations associated to $\delta_B$, $\delta_A$ and $\delta_P$ can be observed,
especially for what concerns $\delta_A$.

The rotorcraft stability is next evaluated through the eigenvalue analysis applied to the linearized motion equations.
The root locus is calculated for various $h/R$ in level flight varying the velocity from 0 to 35 m s$^{-1}$ (see Fig.12 a b c d),
whereas the several modes are recognized by means of the eigenvector analysis.
Fig. 13a shows the root locus out of ground effect.
According to Ref. 1, at low speed the analysis yields two unstable modes which are the phugoid ($Ph$) and the
lateral oscillation ($L.O.$), two aperiodic and stable modes such as the roll ($R.S.$) and pitch ($P.S.$) subsidence,
and  spiral ($Sp$) and heave ($Hv$) modes, that are both stable.
Because of the vehicle axial symmetry about $z_B$, the two modes $Ph$ and $L.O.$ degenerate in a single conical mode whose
eigenvalues collapse as $V$ $\rightarrow$ 0.
As the flight speed increases, the phugoid eigenvalue diminishes its imaginary part
until to intersect the horizontal axis at about 8 m s$^{-1}$,
where two unstable longitudinal aperiodic modes, i.e. A and  B (see Fig.13a), appear.
As for $L.O.$ and $R.S.$, the corresponding eigenvalues are approximatively constant until to a velocity
of about 20 m s$^{-1}$, whereas the pitch subsidence $P.S.$ eigenvalue decreases, reaching its minimum at about 20 m s$^{-1}$.
It is worth to remark that, in the eigenvectors calculated out of ground effect
the Euler angles lag the other state variables and this is in agreement to the classical flight mechanics,
where the Euler angles do not appear in the moment equations of motion.
In ground effect the root locus presents some modifications which become more pronounced as $h/R$ goes to zero.
In particular, now, for $h/R \rightarrow$ 0, the eigenvectors associated to  $Sp$, $P.S.$ and $R.S.$ show Euler angles that
do not lag the other state variables.
This is a peculiarity of all the vehicles that fly in proximity of the ground,
which is caused by the fact that the R.H.S. of the moment motion equations
depend on the Euler angles through the aerodynamic coefficients$^{27}$ and thrust terms.
For an assigned velocity, a significant reduction of the eigenvalues real part is observed for $h/R \rightarrow$ 0.
Because of the hull and rotors moments, under $h/R$ $<$ 2 $ \ Hv$ becomes an oscillating and stable mode whose
eigenvalue imaginary part increases when $h/R$ goes to zero, whereas the phugoid degenerates in a sort of short period mode
whose stability margin increases when the vehicle approaches to the ground.
As for the lateral oscillation, it increases its eigenvalue imaginary part, and results to be stable for $h/R$ $<$ 2.
According to the dynamics of the ground effect machines$^{28, 29}$, the eigenvalues-eigenvectors analysis shows that at relatively low velocities, for $h/R \approx $ 0.5, all the modes result to be stable (See Fig. 13 d), due to the fact that the derivatives ${C_m}_\theta$, ${{C_m}_T}_\theta$,${C_l}_\varphi$ and ${{C_l}_T}_\varphi$ are negative.
Therefore, the Euler angles play a role of  paramount importance in the rotorctaft stability in ground effect,
especially at low $h/R$. 

\bc
{\bf Conclusion}
\ec

This study analyzes the performance and stability of a shrouded-fan UAV by means of
two models which determine force and moment developed by the vehicle in the different flight conditions.
The models calculate force and moment developed by the hull and by the rotors and  
take into account the reciprocal interaction between fuselage and rotors and the
ground effect. These features make the main advantage of the proposed method with respect to
the other models known in the literature.
The limitation of the method is that both the models are based on the potential flow theory where the flow structure
about the airframe is in advance assigned and corresponds to the normal working state.
The present study shows that the two models give results in good agreement with the various source of data existing in the literature.

The trim analysis in and out of ground effect was accomplished and the corresponding results are discussed.
Out of ground effect the same characteristics just known in the literature 
are recovered, whereas in ground effect sizable changing
in the controls that are caused by the interaction between ground and rotorcraft are observed.
The influence of the ground inclination with respect to the horizon is also studied which is responsible of  
seeable changing in the controls especially for what concerns the lateral cyclic pitch.  

Finally the vehicle stability is analyzed through the eigenvalues-eigenvectors analysis of the linearized
motion equations. While the stability out of ground effect is in somewhat comparable with
the data known in the literature, the hull and rotors force and moment, developed in ground effect,
cause great modifications in the vehicle stability.
According to the classical flight mechanics, the modes calculated out of ground effect exhibit
Euler angles which lag the other state variables since such angles do not appear in the R.H.S. of the motion equations,
whereas in ground effect, due to the incremental force and moment which depend on the
Euler angles, the diverse eigenvectors show the Euler angles that do not lag the other state variables.
This property, which yields stable modes in ground proximity, is in agreement with the behavior observed in the ground effect machines.

\bc
{\bf Acknowledgment}
\ec

This work was partially supported by Italian Ministry of University.

\bigskip

\bc
{\bf References}
\ec

\medskip
\no \hspace*{-1.50mm}$^1$ Avanzini G., D'Angelo S., de Matteis G. "Performance and Stability of a Ducted-Fan Uninhabited
Aerial Vehicle" {\it Journal of Aircraft}, Vol. 40, No.1, 2003, pp. 86-93.

\medskip
\no \hspace*{-1.50mm}$^2$ de Divitiis N. "Aerodynamic Modeling and Performance Analysis of a Shrouded Fan Unmanned
Aerial Vehicle".  23th ICAS, {\it International Council of the Aeronautical Sciences}, Paper n.256, Sept.2002.

\medskip
\no \hspace*{-1.50mm}$^{3}$ Newman S. {\it The Foundation of Helicopter Flight}, Ed. Arnold editor, 1994, pp. 49-62.

\medskip
\no \hspace*{-1.50mm}$^{4}$ Leishman J.G., {\it  Principles of Helicopter Aerodynamics},
            Cambridge University Press, 2000.
      
\medskip
\no \hspace*{-1.50mm}$^{5}$ Kohlman, D.  L., {\it  Introduction to V/STOL Airplanes},
           Iowa State Univ. Press, Ames, IA, 1981, pp. 45-81.

\medskip
\no \hspace*{-1.50mm}$^{6}$ Padfield G.D.,
{\it  Helicopter Flight Dynamics: The Theory and Application of Flying Qualities and Simulation Modeling}, AIAA
Education Series, 1996.

\medskip
\no \hspace*{-1.50mm}$^{7}$ McCormick B. W., {\it  Aerodynamics of V/STOL Flight}
           Dover, New York, 1999,  pp. 231-259, 310-321. 
   
\medskip
\no \hspace*{-1.50mm}$^{8}$ Bramwell, A.R.S., Done G., Balmford D.,
{\it  Bramwell's Helicopter Dynamics}, Second edition by Butterworth Heinemann, 2001.

\medskip
\no \hspace*{-1.50mm}$^{9}$ Prouty R.W.,
{\it  Helicopter Performance, Stability, and Control}, Krieger Publishing Company Inc., 1986.

\medskip
\no \hspace*{-1.50mm}$^{10}$ Wyatt, L. A.,
"Static Test of Ground Effect on Planforms Fitted with a Centrally-Located Round Lifting Jet," Ministry of Aviation,
CP-749, June 1962.

\medskip
\no \hspace*{-1.50mm}$^{11}$ Levin B. D., Wardwell D. A. "Single Jet-Induced Effect on Small-Scale Hover Data in Ground Effect"
 {\it Journal of Aircraft}, Vol. 34, No.3, 1997, pp. 400-407.

\medskip
\no \hspace*{-1.50mm}$^{12}$ Xin H., Prasad J.V.R., Peters D.A., Itoga N., Iboshi N., Nagashima T.,
"Ground Effect Aerodynamics of Lifting Rotors Hovering Above Inclined Ground Plane", AIAA-99-3223,
PP. 797-807.

\medskip
\no \hspace*{-1.50mm}$^{13}$ Avanzini G., de Matteis G., "Design of a Shipboard Recovery System for a Shrouded-Fan UAV", 23th ICAS, {\it International Council of the Aeronautical Sciences},  5.6.3., Sept. 2002


\medskip
\no \hspace*{-1.50mm}$^{14}$ Xin H., Prasad J.V.R., Peters D.A.,
"Unsteady Aerodynamics of Helicopter Rotor Hovering in Dynamic Ground Effect", AIAA-98-4456,
PP. 711-720.

\medskip
\no \hspace*{-1.50mm}$^{15}$ Xin H., Prasad J.V.R., Peters D.A.,
"Dynamic Inflow Modeling for Simulation of a Helicopter Operating in Ground Effect", AIAA-99-4114,
PP. 182-191.

\medskip
\no \hspace*{-1.50mm}$^{16}$ Cheeseman I.C., Bennet W.E. "The Effect of the Ground on a Helicopter Rotor in Forward Flight"
Aeronautical Research Council R \& M No. 3021, 1957.

\medskip
\no \hspace*{-1.50mm}$^{17}$ Hayden J.S.,"The Effect of the Ground on Helicopter Hovering Power Required"
32th Annual National V/STOL Forum of the American Helicopter Soc., Washington DC, May 10-12, 1976.

\medskip
\no \hspace*{-1.50mm}$^{18}$ Lamb, H., ``On the Motion of Solids Through a Liquid'', {\it Hydrodynamics},
                6th ed., Dover, New York, 1945, pp. 160-201.

\medskip
\no \hspace*{-1.50mm}$^{19}$Rozhdestvensky, K.V., {\it  Aerodynamics of a Lifting System in Extreme Ground Effect}, Springer-Verlag, 2000,  pp.263-318.

\medskip
\no \hspace*{-1.50mm}$^{20}$ Horlock J. H., {\it  Actuator Disk Theory},
           McGraw-Hill, New York,  1978,  pp. 12-50. 

\medskip
\no \hspace*{-1.50mm}$^{21}$ Etkin, B., {\it  Dynamics of Atmospheric Flight.}
           John Wiley \& Sons, New York, 1972,  pp.~104, 152.

\medskip
\no \hspace*{-1.50mm}$^{22}$ Analytical Methods Inc. VSAERO User's Manual. Revision E5, April, 1994, pp. 1-52.

\medskip
\no \hspace*{-1.50mm}$^{23}$ Kriebel A. R. "Theoretical Stability Derivatives for a Ducted Propeller",
 {\it Journal of Aircraft},  Vol. 1, No.4, 1964, pp. 203-210.

\medskip
\no \hspace*{-1.50mm}$^{24}$ Lamar, J. E.; Margason, R. J."Vortex-lattice FORTRAN program for estimating 
subsonic aerodynamic characteristics of complex planforms", {\it  NASA-TN-D-6142; L-7262}, ; Feb 1, 1971. 
            
\medskip
\no \hspace*{-1.50mm}$^{25}$ Stevens, B.L., and Lewis, F.L. {\it  Aircraft Control and Simulation},
            John Wiley \& Sons, New York, 1992.

 \medskip
\no \hspace*{-1.50mm}$^{26}$ Cycon J.P., "Sikorsky Aircraft UAV Program", {\it Verflite}, Vol. 38, No. 3, 1992, pp. 26-30.

\medskip
\no \hspace*{-1.50mm}$^{27}$ de Divitiis N., "Performance and Stability of a Winged Vehicle in Ground Effect",
 {\it Journal of Aircraft},  Vol. 42, No.1, 2005, pp. 148-157.

\medskip
\no \hspace*{-1.50mm}$^{28}$ Walker N. K. "Influence of Fan and Ducting Characteristics on the Stability of Ground Effect Machines",
 {\it Journal of Aircraft},  Vol. 2, No.1, 1965, pp. 25-32.

\medskip
\no \hspace*{-1.50mm}$^{29}$ Campbell, J.P., {\it  Vertical Takeoff and Landing Aircraft.}
           MacMillan, New York, 1962,  pp.~106, 171.









\newpage

\begin{center}
{ Table 1. Characteristics of the UAV.}
\end{center}
\begin{center}
 \begin{tabular}{lr}
Overall diameter ($m$)            & 1.9   \\[1pt]
Rotor diameter ($m$)              &   1.1 \\
Central hub diameter ($m$)        &  0.25 \\
Maximum Overall weight  ($N$)    & 800   \\
Payload                   ($N$)   & 100   \\
Coaxial rotors                    & 2     \\
Power              ($h.p.$)       & 3 $\times$ $14$ \\
at                  ($RPM$)       & $11000$ \\
Rotor speed   ($RPM$)             & 3000 \\
Endurance ($h$)                   & 1.5  \\
Service ceiling ($m$)             & 2000
\end{tabular}
\end{center}

\newpage

\centerline { \bf List of captions }

\bigskip
\bigskip
\bigskip
\bigskip

\centerline {Fig. 1 Reference rotorcraft.}

\bigskip
\bigskip
\bigskip
\bigskip

\centerline {Fig. 2 Rotor working regimes.}

\bigskip
\bigskip
\bigskip
\bigskip

\centerline {Fig. 3 Vehicle aerodynamics in normal working state out of ground effect.}

\bigskip
\bigskip
\bigskip
\bigskip

\centerline {Fig. 4 Vehicle aerodynamics in ground effect.}

\bigskip
\bigskip
\bigskip
\bigskip

\centerline {Fig. 5  Aerodynamic coefficients out of ground effect.}

\bigskip
\bigskip
\bigskip
\bigskip

\centerline {Fig. 6  Aerodynamic coefficients out of ground effect: }
\centerline {Continuous lines for the present data. The symbols are from Ref. 23.  }

\bigskip
\bigskip
\bigskip
\bigskip

\centerline {Fig. 7  Aerodynamic coefficients in ground effect.}

\bigskip
\bigskip
\bigskip
\bigskip

\centerline {Fig. 8  Thrust and torque coefficients in axial flight out of ground effect:}
\centerline {dashed and continuous lines are, respectively, for free and shrouded rotors.}

\bigskip
\bigskip
\bigskip
\bigskip

\centerline {Fig. 9 Rotor in ground effect.}

\bigskip
\bigskip
\bigskip
\bigskip

\centerline {Fig. 10 Influence of height, attitude and sink rate}
\centerline {on the rotors characteristics.}

\bigskip
\bigskip
\bigskip
\bigskip

\centerline {Fig. 11 Trim calculation in horizontal flight at different $h/R$.}

\bigskip
\bigskip
\bigskip
\bigskip

\centerline {Fig. 12 Trim controls at $h/R$ =1 with inclined ground wall.}

\bigskip
\bigskip
\bigskip
\bigskip

\centerline {Fig. 13 Root Locus in horizontal flight at various $h/R$.}

\newpage

\begin{figure}[!ht]
\begin{center}
\psfig{file=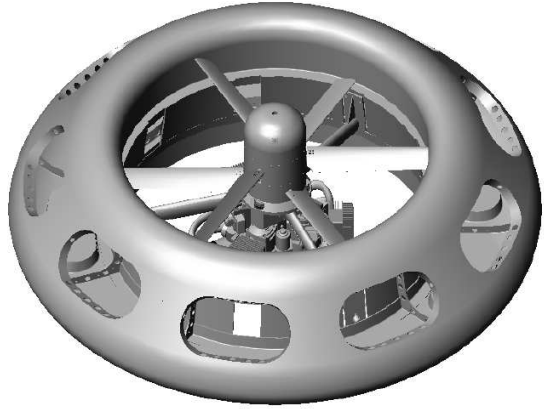,width=11.cm}
\caption{Reference rotorcraft.}
\end{center}
\end{figure}

\newpage

\begin{figure}[!ht]
\begin{center}
\psfig{file=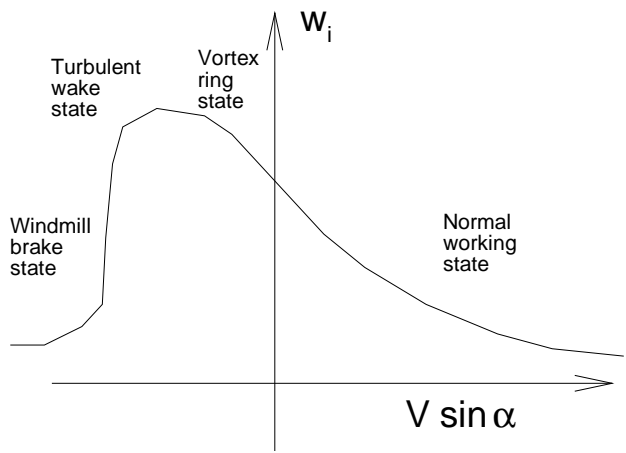,width=8.cm}
\vspace{0.mm}
\caption{Rotor working regimes.}
\end{center}
\end{figure}

\newpage

\begin{figure}[!ht]
\begin{center}
\psfig{file=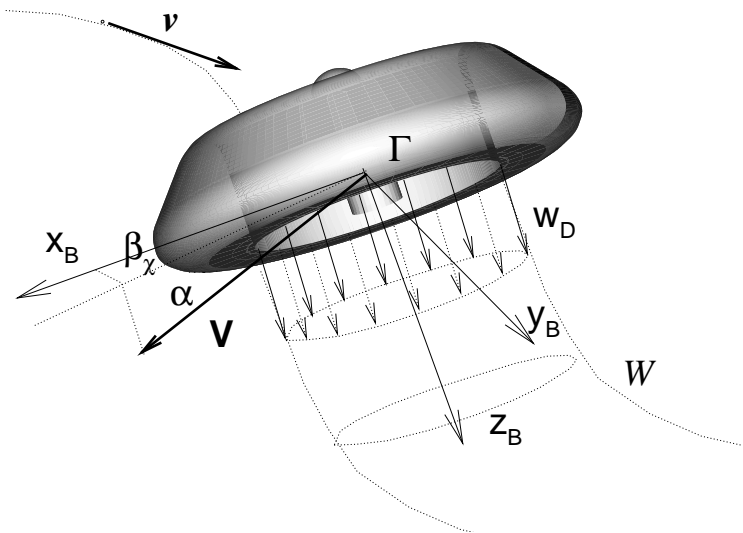,width=8.cm}
\vspace{0.mm}
\caption{Vehicle aerodynamics in normal working state out of ground effect.}
\end{center}
\end{figure}

\newpage

\begin{figure}[!ht]
\begin{center}
\psfig{file=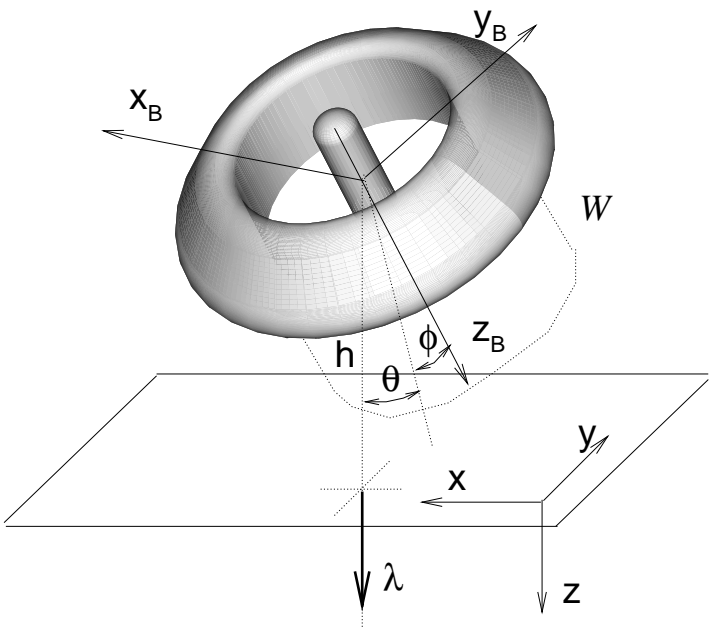,rheight=0.0cm}
\vspace{0.mm}
\caption{Vehicle aerodynamics in ground effect.}
\end{center}
\end{figure}

\newpage

\begin{figure}[!ht]
\begin{center}
\psfig{file=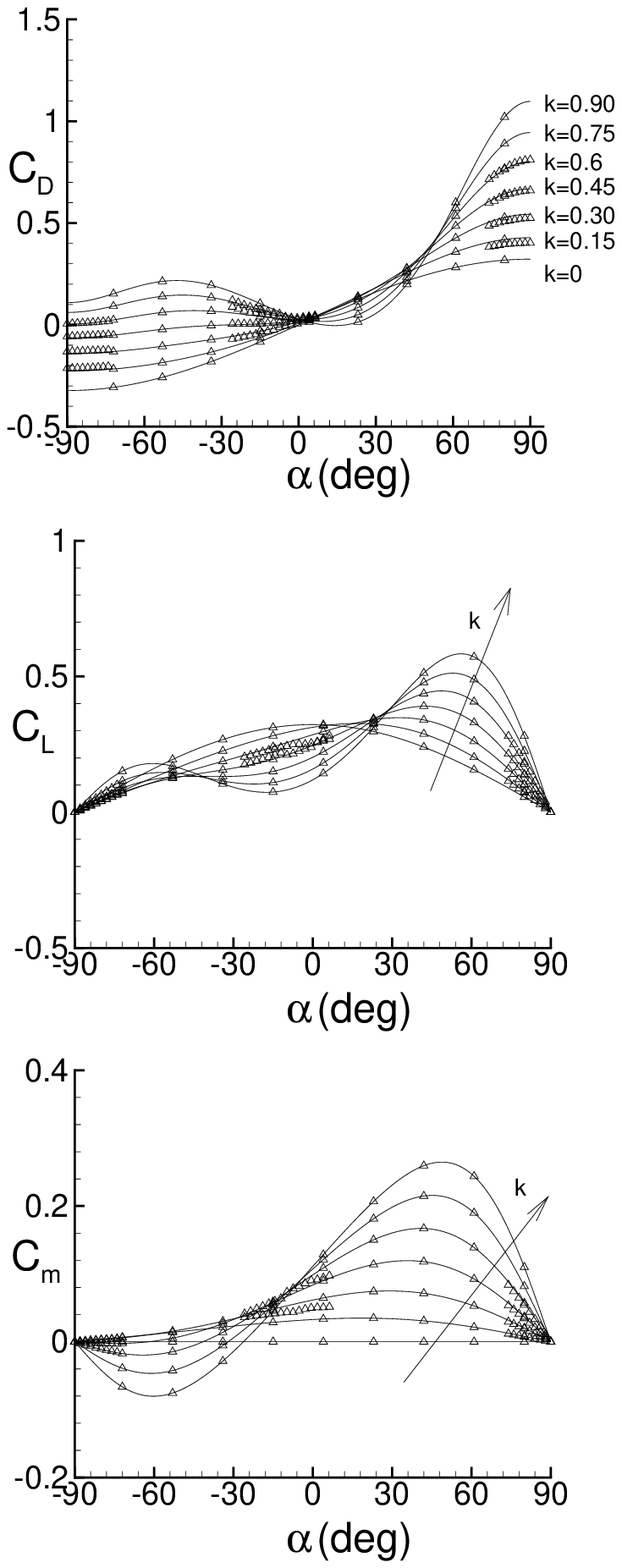,rheight=0.0cm}
\vspace{0.mm}
\caption{Aerodynamic coefficients out of ground effect.}
\end{center}
\end{figure}

\newpage

\begin{figure}[!ht]
\begin{center}
\psfig{file=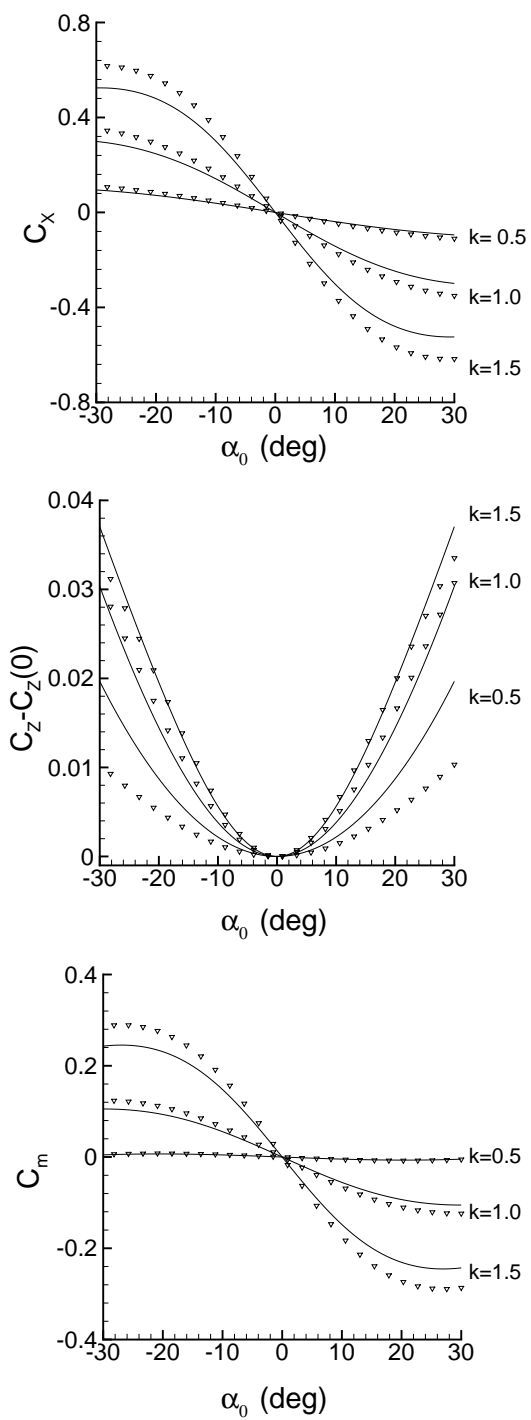,rheight=0.0cm}
\vspace{0.mm}
\caption{ Aerodynamic coefficients out of ground effect: Continuous lines for the present data. The symbols are from Ref. 23.  }
\end{center}
\end{figure}

\newpage

\begin{figure}[!ht]
\begin{center}
\psfig{file=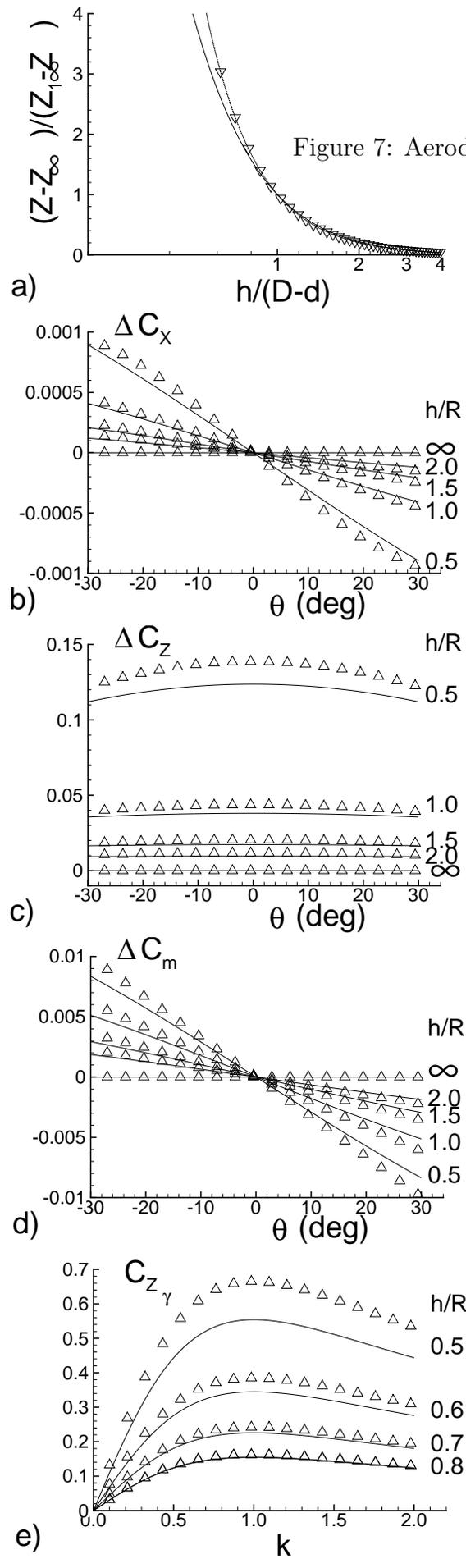,rheight=0.0cm}
\vspace{0.mm}
\caption{Aerodynamic coefficients in ground effect.}
\end{center}
\end{figure}

\newpage

\begin{figure}[!ht]
\begin{center}
\psfig{file=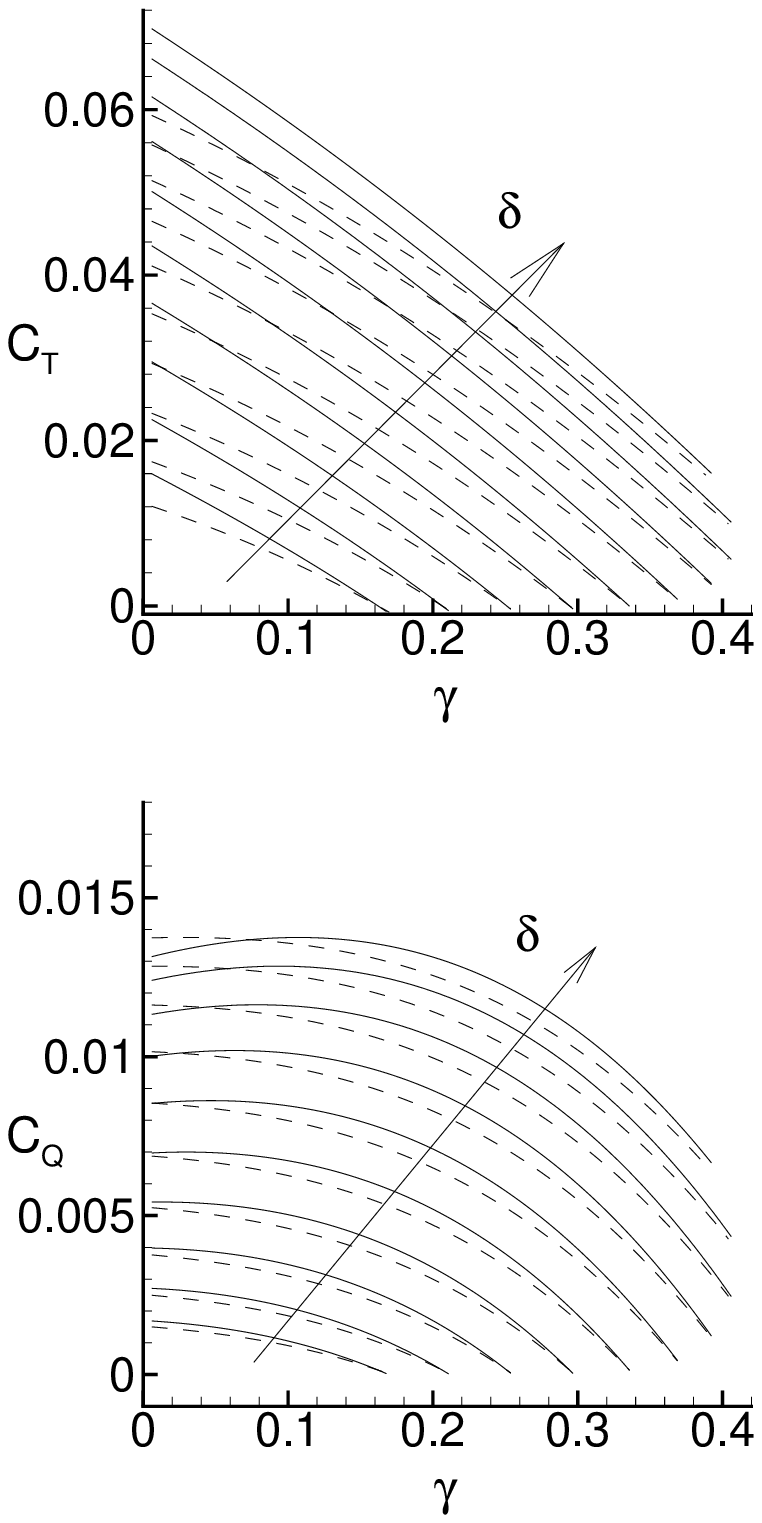,rheight=0.0cm}
\vspace{0.mm}
\caption{Thrust and torque coefficients in axial flight out of ground effect: dashed and continuous lines are, respectively, for free and shrouded rotors.}
\end{center}
\end{figure}

\newpage

\begin{figure}[!ht]
\begin{center}
\psfig{file=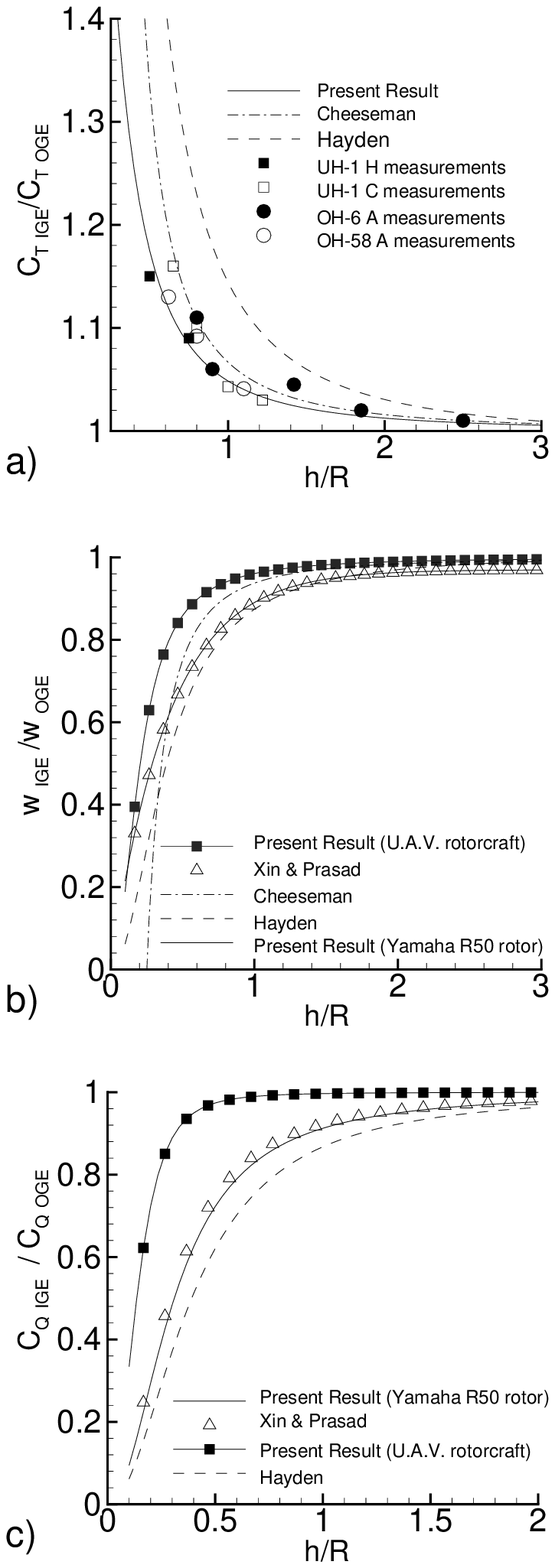,rheight=0.0cm}
\vspace{0.mm}
\caption{Rotor in ground effect.}
\end{center}
\end{figure}

\newpage

\begin{figure}[!ht]
\begin{center}
\psfig{file=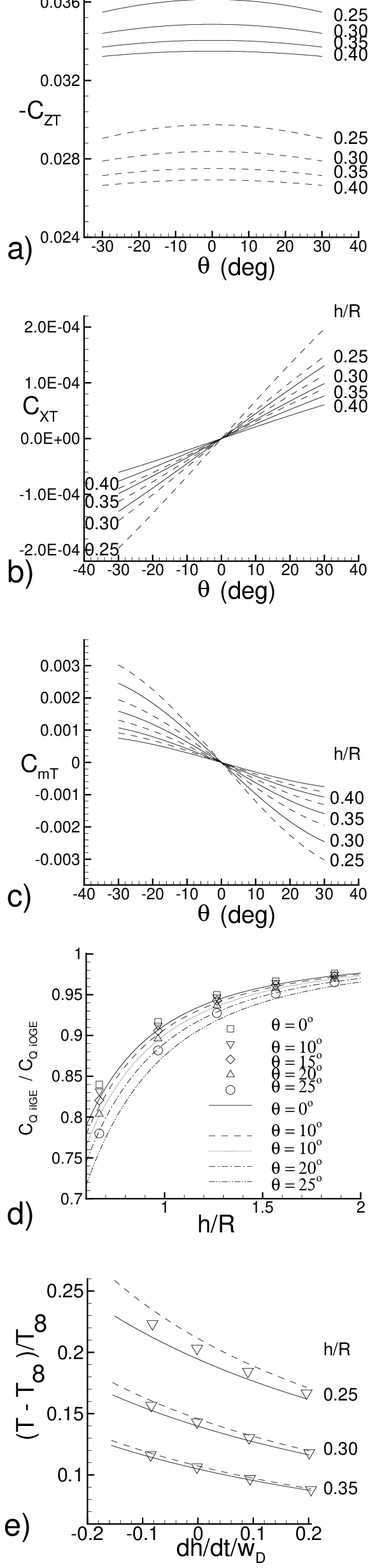,rheight=0.0cm}
\vspace{-18.mm}
\caption{Influence of height, attitude and sink rate on the rotors characteristics.}
\end{center}
\end{figure}

\newpage

\begin{figure}[!ht]
\begin{center}
\psfig{file=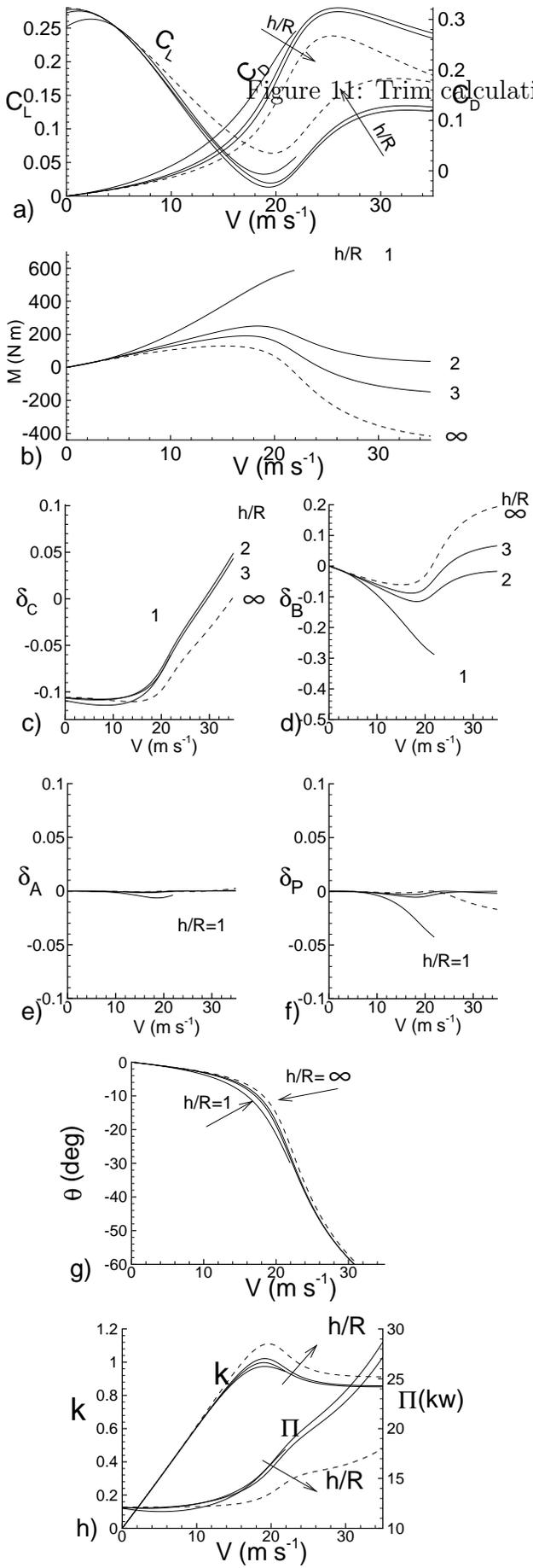,rheight=0.0cm}
\vspace{0.mm}
\caption{Trim calculation in horizontal flight at different $h/R$.}
\end{center}
\end{figure}

\newpage

\begin{figure}[!ht]
\begin{center}
\psfig{file=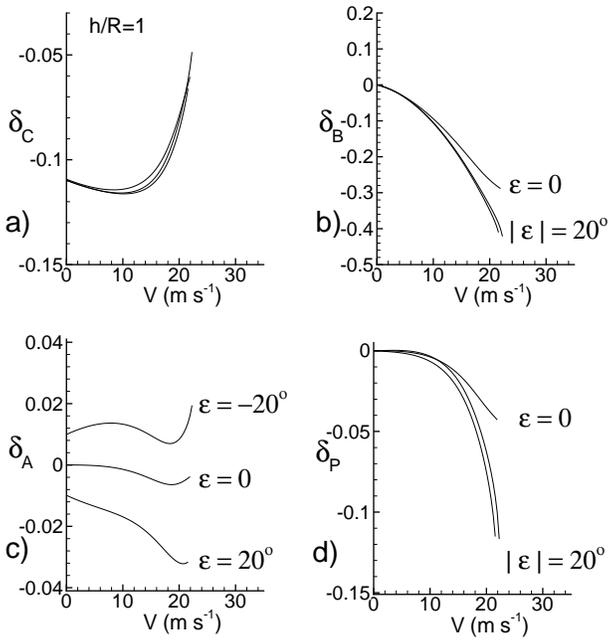,rheight=0.0cm}
\vspace{0.mm}
\caption{Trim controls at $h/R$ =1 with inclined ground wall.}
\end{center}
\end{figure}

\newpage

\begin{figure}[!ht]
\begin{center}
\psfig{file=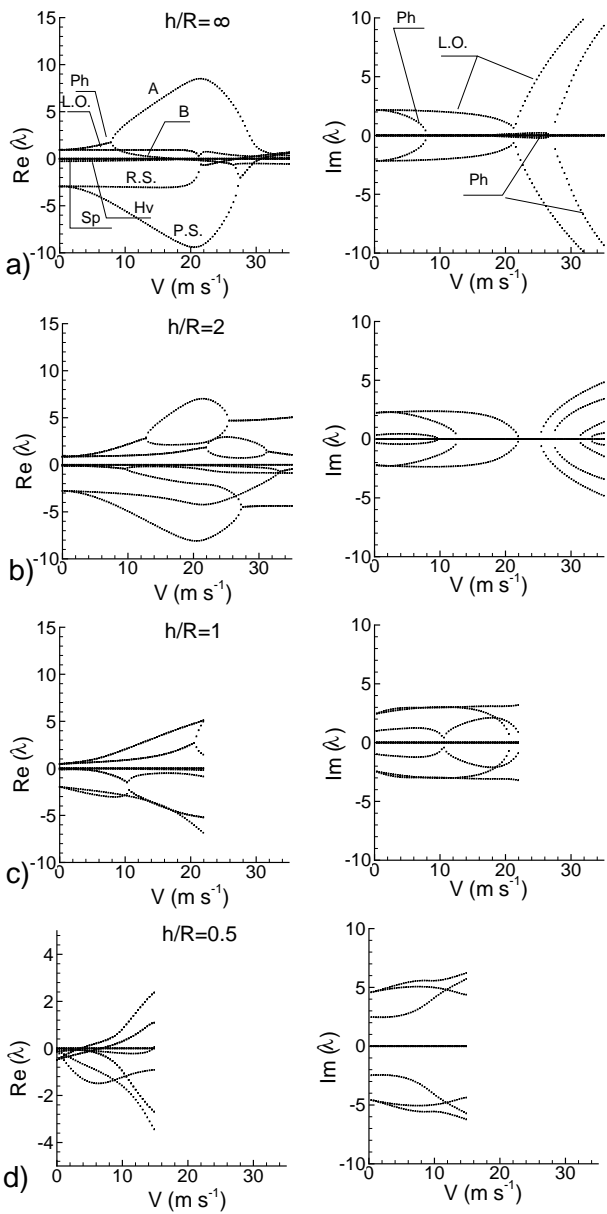,rheight=0.0cm}
\vspace{-0.mm}
\caption{Root Locus in horizontal flight at various $h/R$.}
\end{center}
\end{figure}

\end{document}